\documentclass{PoS}

\usepackage{graphicx}
\usepackage[tight,footnotesize]{subfigure}
\usepackage{amsmath}

\title{Mesonic String of Diquark-Quark Configuration at Finite Temperature}
\ShortTitle{Mesonic String of Diquark-Quark}
\author{\speaker{A. Bakry}, X. Chen, M. Deliyergiyev, A. Galal, S. Xu and P.M.
 Zhang\thanks{
 	This work has been funded by the Chinese Academy of Sciences President's International Fellowship Initiative grants No.2015PM062 and No.2016PM043, the Recruitment Program of Foreign Experts, NSFC grants (Nos. 11035006, 11175215, 11175220) and the Hundred Talent Program of the Chinese Academy of Sciences (Y101020BR0).
 	}\\
 
 Institute of Modern Physics, Chinese Academy of Sciences, Gansu 730000, China\\
 E-mail: \email{ahmed.bakry@mail.com}}
\abstract{
	 We investigate the distance and temperatures scale for which the string in baryonic quark configuration approaches the limiting behavior of mesonic strings in pure Yang-Mills SU(3) lattice gauge theory. We calculate and compare the numerical values of the Polyakov loop correlators and the width profile of both diquark-quark $(QQ)Q$ and mesonic $Q\bar{Q}$ strings. We find the diquark-quark configuration to exhibit an identical behavior to the mesonic string for the potential and energy-density width profile for temperature near the end of QCD plateau. In the vicinity of the deconfinement point; however, the symmetry in the energy-width profile with the meson does not manifest at both short and intermediate distance scales. Moreover, we observe significantly different numerical values for Polyakov loop correlators corresponding to each system. The splitting of the two identical system suggest that, in the high temperature region of the confined phase of QCD, the subsisted baryonic decouplet states overlap with the excited mesonic spectrum yielding the diquark-quark symmetry with the meson inexact in a small enough neighborhood of the critical point $T_{c}$.}
\FullConference{Preprint submitted to XVII International Conference on Hadron Spectroscopy and Structure - Hadron2017, 25-29 September 2017, University of Salamanca, Salamanca, Spain}

\begin{document}

\section{Numerical Lattice Results}
  We briefly report some of the common properties of gluonic field of the quark-antiquark $Q\bar{Q}$ and baryonic diquark quark $(QQ)Q$ systems. Lattice measurements are taken on a set of $SU(3)$ pure gauge configurations after $n_{sw}$ of an over improved stout-link smearing sweeping. The configurations are generated using the standard Wilson gauge action and heatbath updating on two lattices of a spatial volume of $36^{3}$ ($\beta = 6.00$, lattice spacing $a=0.1$ fm) and temporal extents of $N_{t} = 8$ and $N_{t} = 10$ corresponding to temperatures $T/T_{c} = 0.9$ and $T/T_{c} = 0.8$, respectively~\cite{Bakry:2017utr}. 

  A dimensionless scalar field characterizing the gluonic field can be defined as
\begin{equation}
\begin{split}
\mathcal{C}(\vec{\rho};\vec{r}_{1},\vec{r}_{2}) & = \frac{ \langle\mathcal{P}_{Q\bar{Q}}(\vec{r}_{1},\vec{r}_{2}) S(\vec{\rho})\rangle} {\langle \mathcal{P}_{Q\bar{Q}}(\vec{r}_{1},\vec{r}_{2})\rangle \langle S(\vec{\rho}) \rangle},\\
\mathcal{C}(\vec{\rho};\vec{r}_{1},\vec{r}_{2},\vec{r}_{3}) & = \frac{\langle\mathcal{P}_{3Q}(\vec{r}_{1},\vec{r}_{2},\vec{r}_{3}) S(\vec{\rho})\rangle } {\langle \mathcal{P}_{3Q}(\vec{r}_{1},\vec{r}_{2},\vec{r}_{3})\rangle \langle S(\vec{\rho}) \rangle}
\label{eq:ScalaField}
\end{split}
\end{equation}

\begin{figure*}[th]
\begin{center}
\subfigure[Meson $Q\bar{Q}$]{	
\includegraphics[scale=0.2]{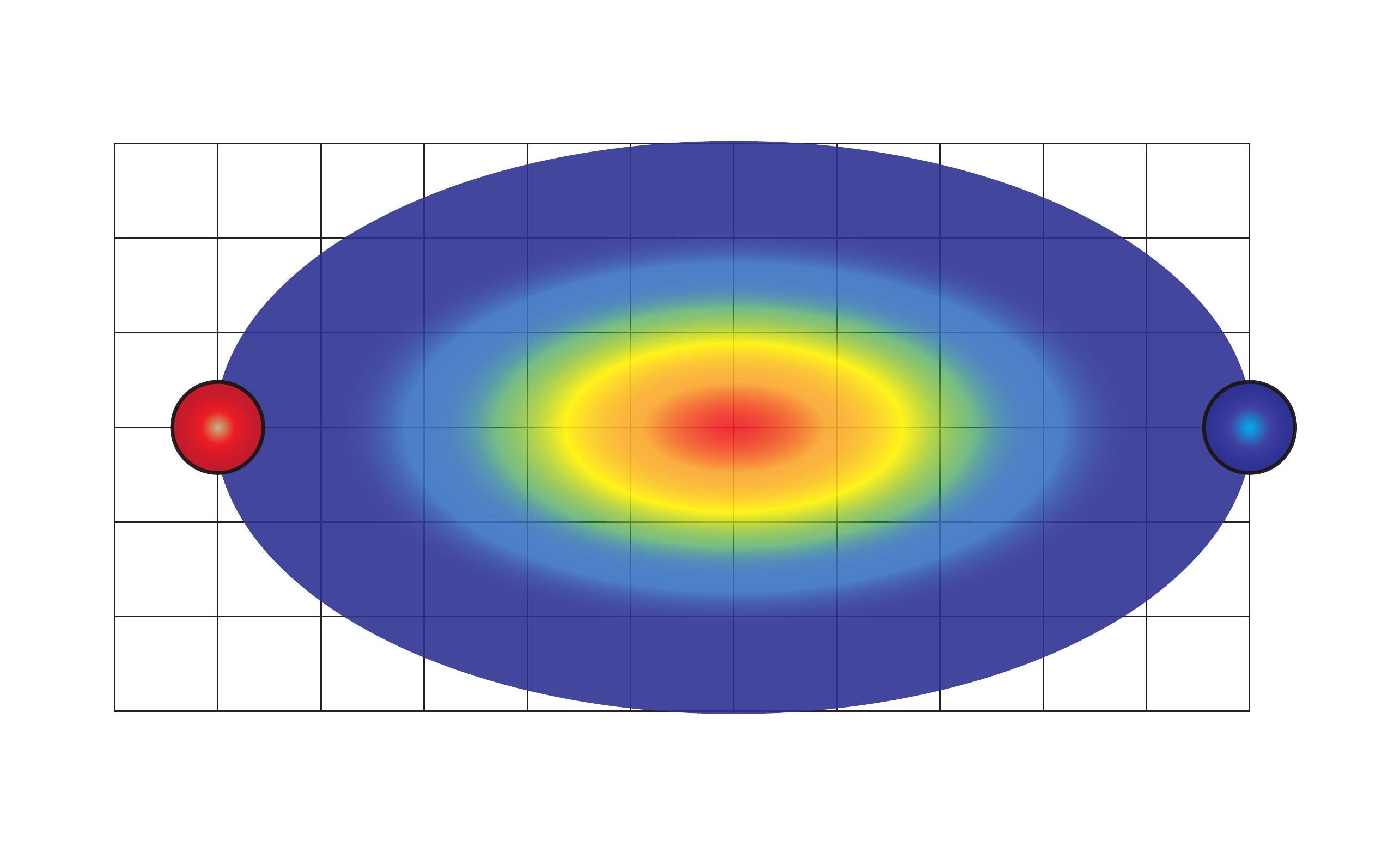}
}
\subfigure[Baryon $(QQ)Q$; base A=0.2 fm]{
\includegraphics[scale=0.2]{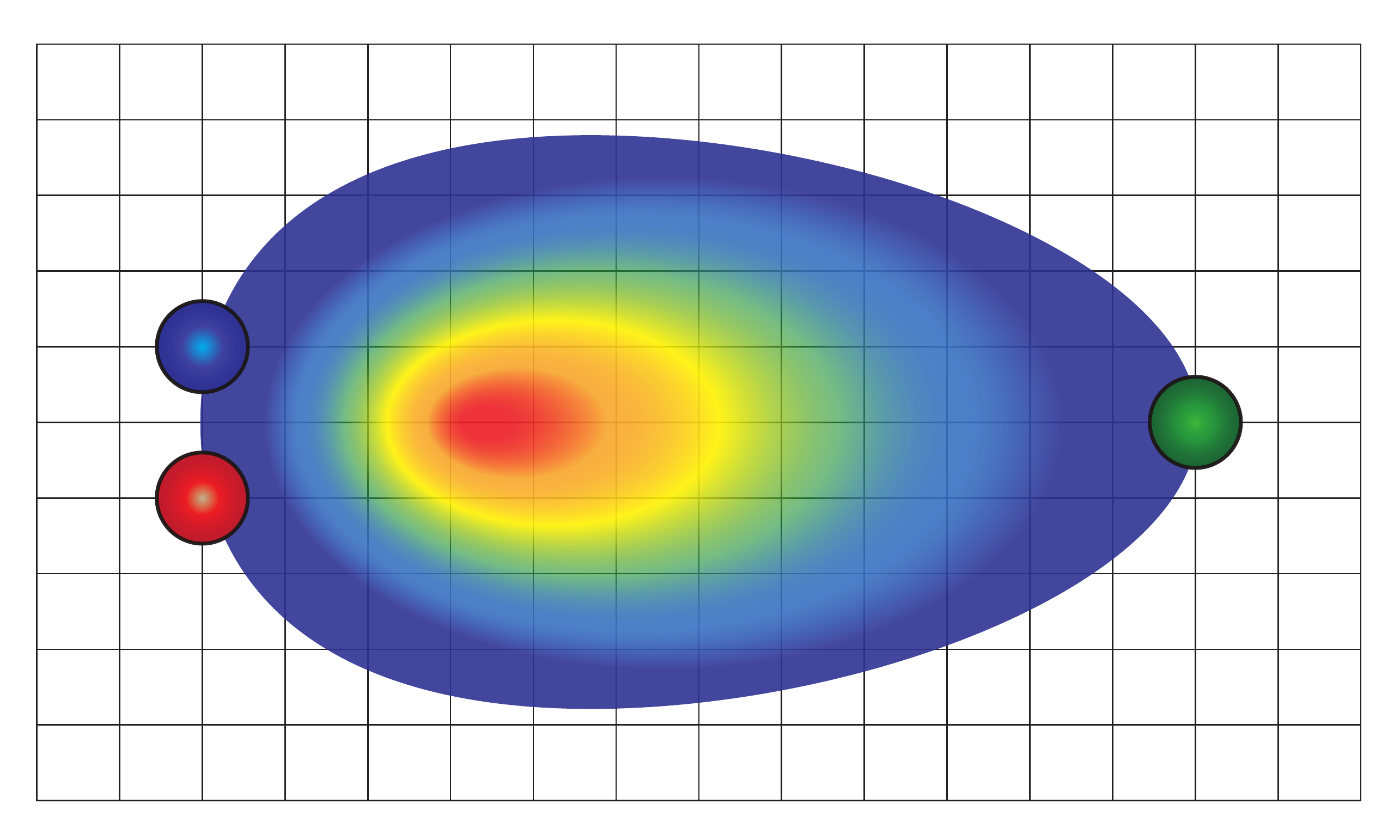}
}	
\subfigure[]{
\includegraphics[scale=0.6]{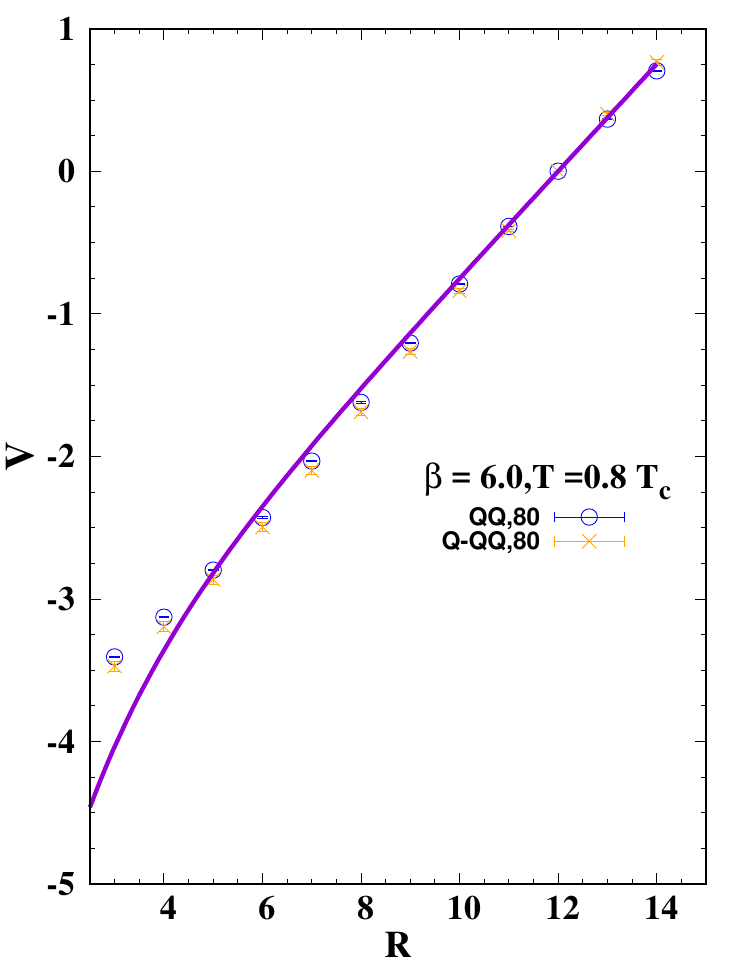}
}
\subfigure[]{
\includegraphics[scale=0.42]{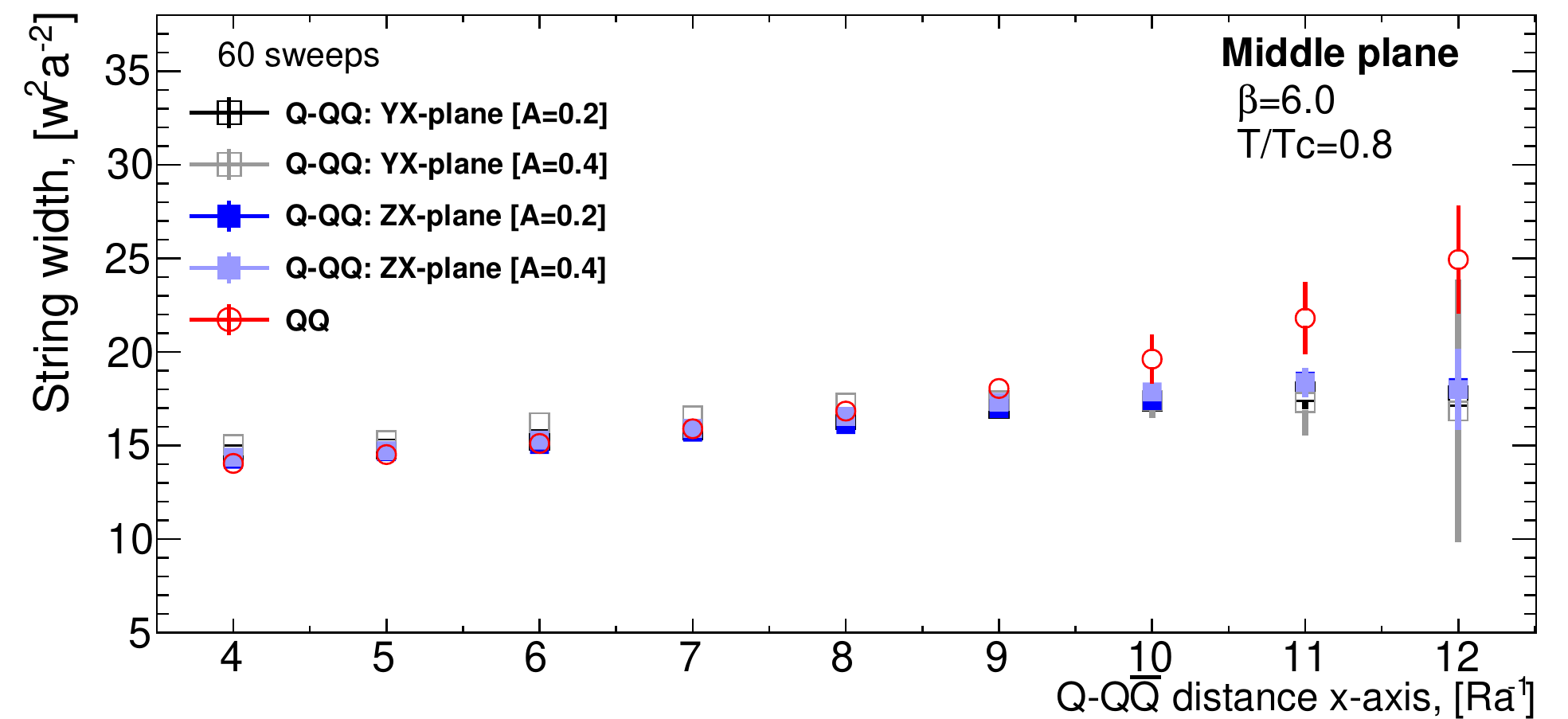}
}

\caption{ The action density Eq.\eqref{eq:ScalaField} for both meson (a) and baryon (b) at $T/T_{c}=0.8$ and height $R=1.0$ fm; for $n_{sw}=60$. 
	(c) The potential of the meson and baryonic systems-lattice units-of base $A=0.2$ fm, line correspond to free string potential~\cite{Bakry:2017utr}.
	(d) The width of the string in the middle plane for $Q\bar{Q}$  and  $(QQ)Q$ of isosceles base $A=0.2,0.4$ fm ``in-plane and off-plane''. 
	} 
\end{center}
\end{figure*}
  for baryonic and meson systems~\cite{Bakry:2017utr}, respectively, with the color charge position $\vec{r}_{i}$ and flux probe vector $\vec{\rho}$.  The $Q\bar{Q}$ and $(QQ)Q$ static potential is extracted from the logarithm~\cite{Bakry:2017utr} of Polyakov loop correlators $\mathcal{P}_{2Q}(\vec{r}_{1}, \vec{r}_{2})$ and $\mathcal{P}_{3Q}(\vec{r}_{1}, \vec{r}_{2}, \vec{r}_{3})$, respectively. The Diquark-quark ($Q\bar{QQ}$) is constructed as an isoscele triangle of two quark at the base (diquark) separated by two lattice spacing.

\begin{figure*}
\begin{center}
\subfigure[]{\includegraphics[scale=0.16]{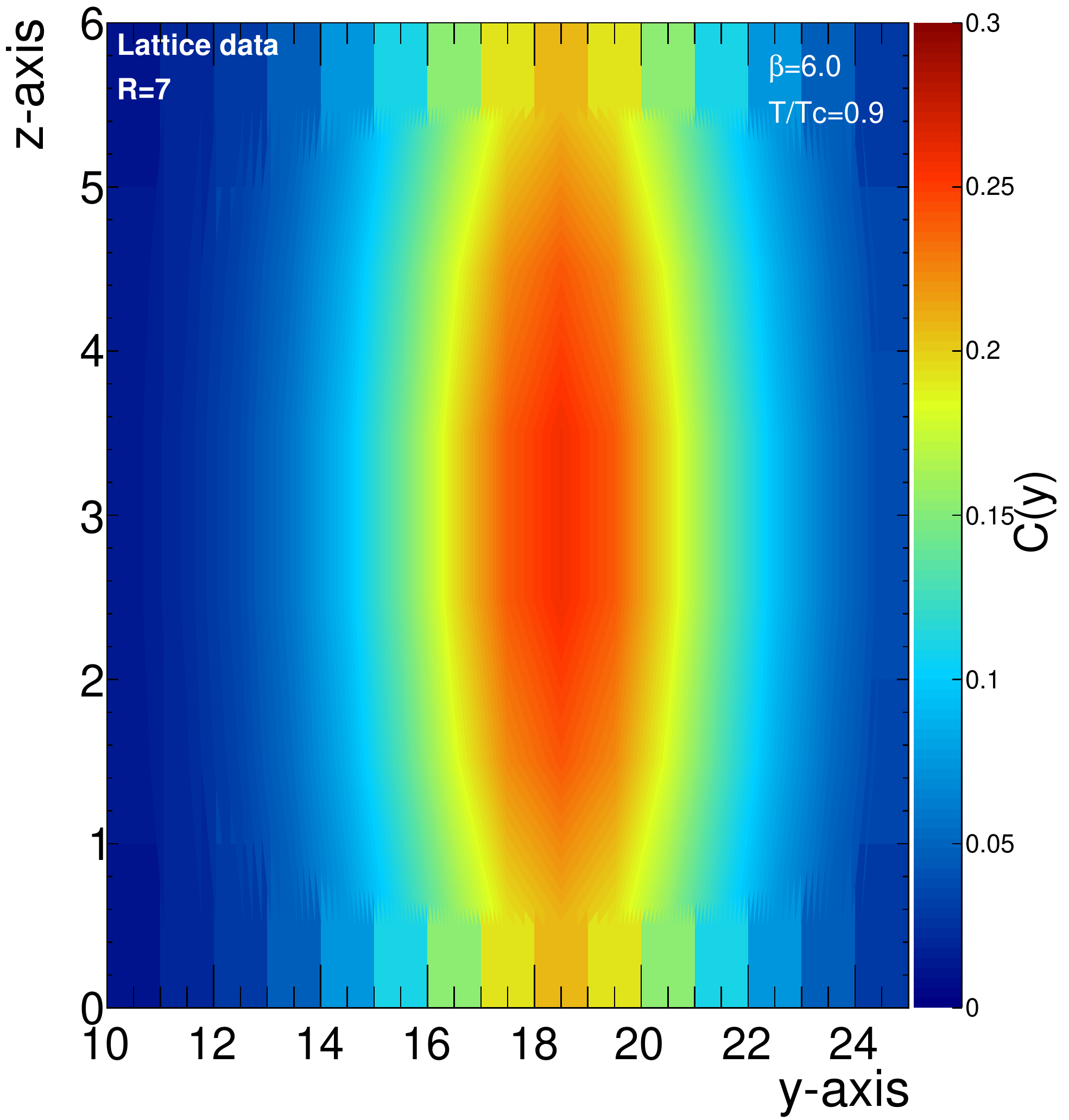}}
\subfigure[]{\includegraphics[scale=0.16]{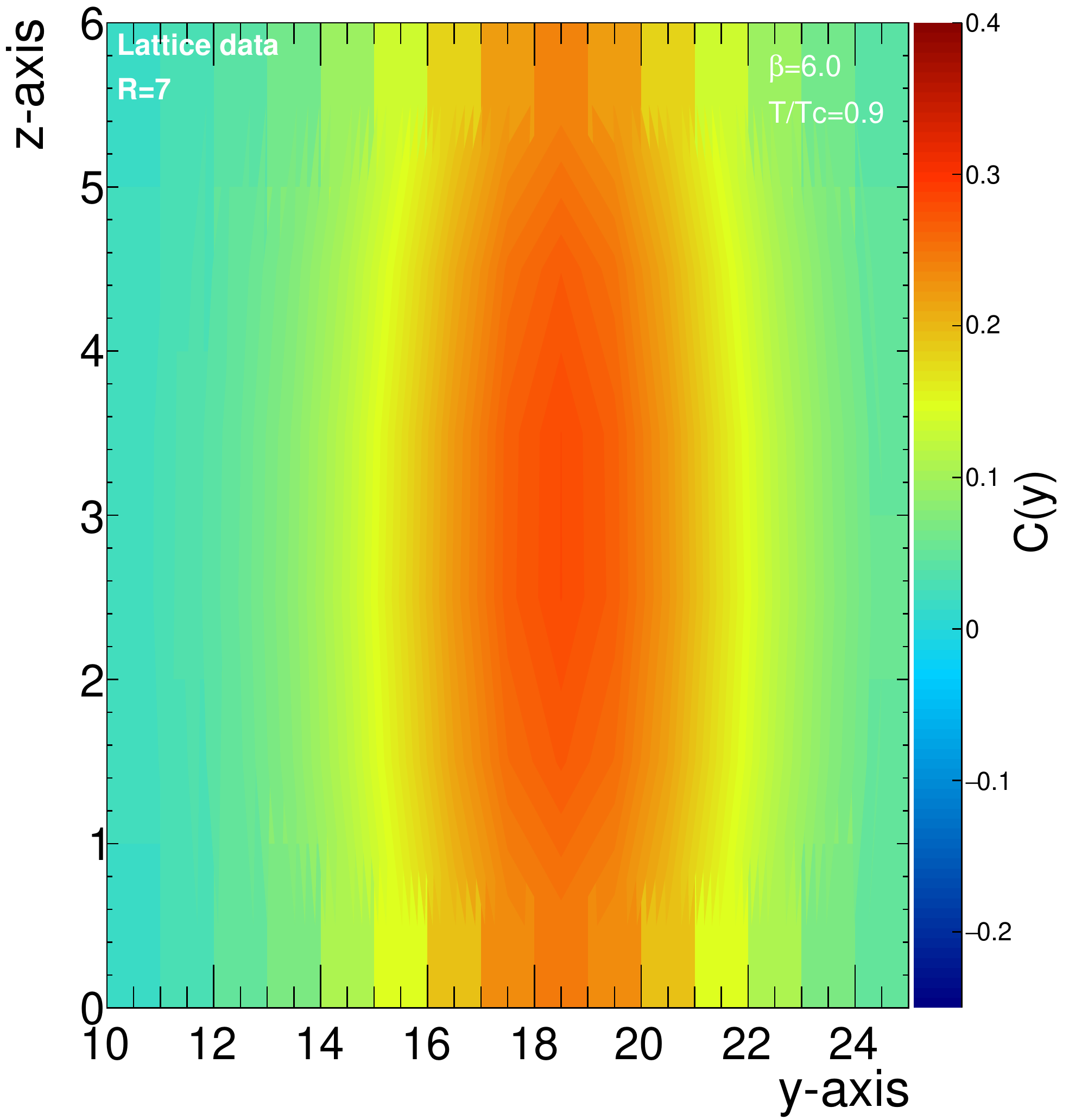}}
\subfigure[]{\includegraphics[scale=0.16]{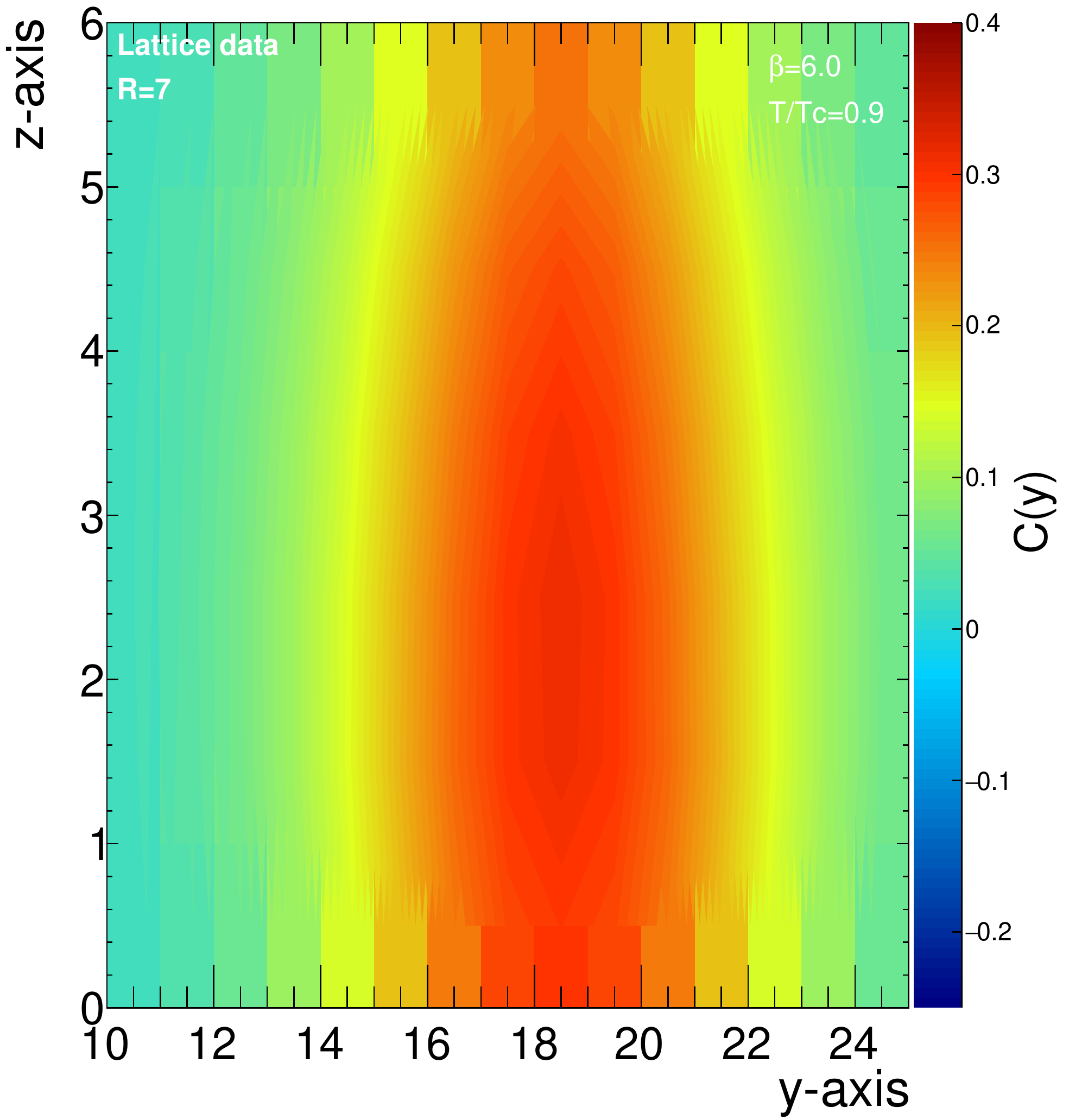}}
\subfigure[]{\includegraphics[scale=0.36]{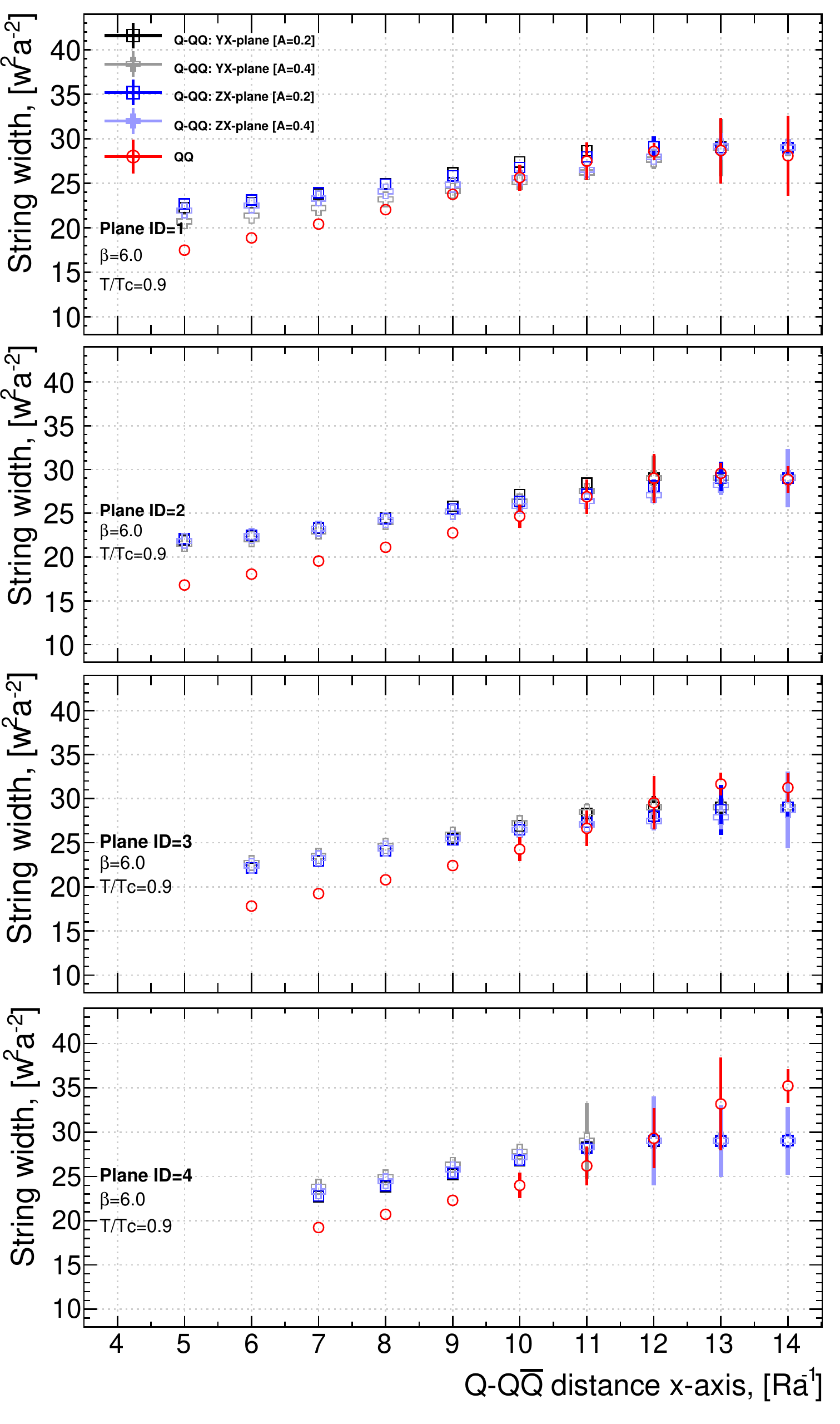}}
\subfigure[]{\includegraphics[scale=0.9]{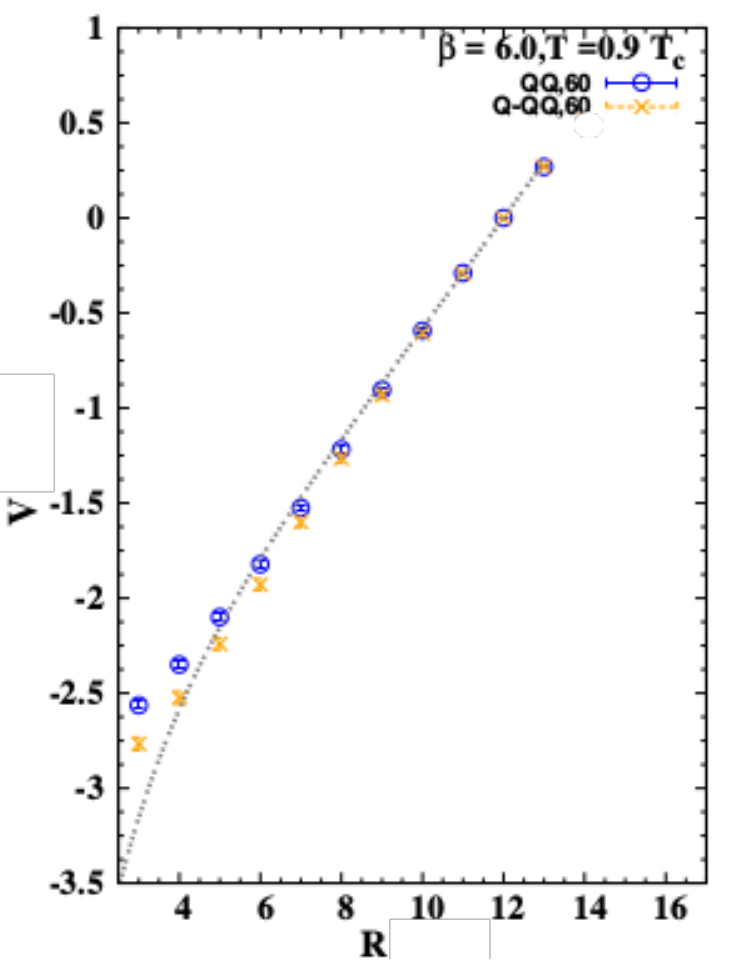}}
\caption{ The in-plane action density Eq.\eqref{eq:ScalaField} for meson $Q\bar{Q}$ in (a) and baryon $(QQ)Q$ with isosceles height R=0.7 fm and base length A=0.2 fm in (b) and A=0.4 fm in (c) at $T/T_{c}=0.9$ and $R=10$ fm, $n_{sw}=60$. (c)The in-plane and off-plane width at $T/T_{c}=0.9$ for $(QQ)Q$ and $Q\bar{Q}$ systems at Planes $y=1,2,3$ and $x=4$, $n_{sw}=40$. (d)The potential of the meson and baryonic systems of base $A=0.2$ fm, $n_{sw}=60$.}   
\end{center}
\end{figure*}

  Generally, we find the energy profiles of $(QQ)Q$ to be very similar considering two different isosceles bases A=0.2, 0.4 fm at any temperature. Moreover, we observe no significant effect of the number of UV filtering sweeps $n_{sw}$ on the observed symmetry or asymmetry between both systems. 

   Near the end of QCD plateau, $T/T_{c}=0.8$ (See Fig.1), in the vicinity of the quark, the action density exhibits cylindrical symmetry with no measurable difference between the transverse profiles of the diquark-quark flux-tubes and the quark-antiquark flux-tubes (See for comparison Ref.~\cite{Bakry:2014gea}). We find that the string tension of the diquark-quark to be the same as the mesonic string~\cite{Bakry:2016aod}. These results are compatible with the recent findings obtained employing the three Polyakov loop analysis at T=0~\cite{Koma:2017hcm} (This instance elaborate on the close analogy between the analysis at end of QCD plateau ~\cite{Bakry:2014gea,Bakry:2016aod} and that at T=0) and that using  the Wilson loop overlap formalism~\cite{Bissey:2009gw}.
  
  In the vicinity of the deconfinement point $T/T_{c}=0.9$ (See Fig.2), the flux density corresponding to each system show very similar profiles near the quark. However, the vacuum expulsion close to the diquark is stronger than that in the proximity of the antiquark. The off-plane and in-plane width of the Q(QQ) system exhibit cylindrical symmetry; even so, the string profile is apparently not identical to the $Q\bar{Q}$ system. The coincidence with the mesonic string does not manifest neither at small nor intermediate separation regions indicating that the string is energetic enough to induce splitting towards a baryonic-like structure~\cite{Bakry:2014gea,Bakry:2016aod}. This is consistent with a significant change in the numerical values of the Polyakov correlators of  $(QQ)Q$ system for $R< 1$ fm with the increase of the temperature as depicted in Fig.~2 (d).
  
  This suggests a continuum constraint on the mesonic limit excluding small enough neighborhoods of $T_{C}$: $\forall \epsilon > 0 $ with $\vert 1-T/T_{c}\vert < \epsilon$, $\exists (R,T)$ such that $\vert V_{Q(QQ)}(R,T)-V_{Q\bar{Q}}(R,T)\vert \textgreater \delta$. 
  
\section{Conclusion and prospect}  

 In this work, we investigate the symmetry between the gluon flux-tubes for the quark-antiquark $Q\bar{Q}$ and three quark systems at finite temperature. We approximate the baryonic quark-diquark $(QQ)Q$ configuration by constructing the two quark at small separation distance of at least 0.2 fm.

  For temperatures up to end of QCD plateau we measure identical mean-square width profile at all separation distance together with close to identical values of Polyakov loop correlators. However, this symmetry between $(QQ)Q$ and $Q\bar{Q}$ systems no longer manifests close to the deconfinement point. The action density and potential of $(QQ)Q$ evidently divulge variant profiles.

  These findings limit the validity~\cite{Koma:2017hcm, Bissey:2009gw} of the expectation that the diquark precisely share many properties in common with the antiquark to the plateau region of QCD; otherwise, excited baryonic states can manifest in the hyperfine structure in small neighborhoods of the QCD critical point. The propagation in the Euclidean time of Wilson loop in addition to the APE link-blocking in ~\cite{Bissey:2009gw} has been sufficient for the excited decuplet baryonic states to decay yielding an optimal overlap with the mesonic ground state, which is evidently not the case at very high temperatures.

  It would be interesting to reiterate these calculations ~\cite{Bakry:2017utr,Bakry:2014gea,Bakry:2016aod} with the employment of finer lattices at lower or higher temperatures or by considering dynamical quarks. One would question also the persistence of meson-baryon symmetries in the presence of strong magnetic fields. This analysis is likely to be of a considerable interest to phenomenological models of hadron structure and will be discussed in a future work.

\end{document}